\documentclass[aps,prl,twocolumn,groupedaddress,showpacs]{revtex4}

\usepackage{graphicx}
\usepackage{amssymb}

\begin{document}

\title{Acoustic Analogue of Bloch Oscillations and Resonant Zener Tunneling in Ultrasonic Superlattices}
\author{Helios Sanchis-Alepuz}
\affiliation{Wave Phenomena Group, Nanophotonics Technology Center, Polytechnic University of Valencia, C/ Camino de Vera s/n, E-46022 Valencia, Spain.}
\author{Yuriy A. Kosevich}
\author{Jos\'e S\'anchez-Dehesa}
\email[E-mail:]{jsdehesa@upvnet.upv.es}
\affiliation{Wave Phenomena Group, Nanophotonics Technology Center, Polytechnic University of Valencia, C/ Camino de Vera s/n, E-46022 Valencia, Spain.}
\date{\today}

\begin{abstract}
We demonstrate the existence of Bloch oscillations of acoustic fields in sound propagation through a superlattice of water cavities and layers of methyl methacrylate. To obtain the acoustic equivalent of a Wannier-Stark ladder, we have employed a set of cavities with different thicknesses. Bloch oscillations are observed as time-resolved oscillations of transmission in a direct analogy to electronic Bloch oscillations in biased semiconductor superlattices. Moreover, for a particular gradient of cavity thicknesses, an overlap of two acoustic minibands occurs, which results in resonant Zener-like transmission enhancement.
\end{abstract}

\maketitle



%


\bigskip

Bloch oscillations (BO) and Zener tunneling are fundamental transport effects appearing when electrons in a periodic potential are accelerated by an external DC electric field\cite{bloch,zener}. Both phenomena have been impressively demonstrated in a number of experiments after the advent of semiconductor superlattices (SSL)\cite{esaki}. The frequency domain counterpart of BO is the equidistant Wannier-Stark ladders (WSL) 
of the electronic states in a biased superlattice, leading to resonances of the density of states which were observed in optical spectra\cite{mendez}. In time-resolved optical experiments, BO were first observed as oscillations of electron wave packets in biased SSL \cite{feldmann,leo,wash,dekorsy,loeser}, and later as a periodic motion of ensembles of ultracold atoms\cite{bendahan,wilkinson} and Bose-Einstein condensates\cite{anderson,morsch}. The related high-field phenomena, the nonresonant Zener tunneling 
between neighboring minibands and resonant Zener-like tunneling between the anticrossing Wannier-Stark states of neighboring minibands,  
were also observed in SSL\cite{schneider,rosam}. 

In general, the Bragg reflection can cause BO of a wave of any nature (electronic, optical, acoustical or matter wave) in a lattice with a weak linear gradient of the lattice potential, which can be caused in turn by an external field or perturbation of any nature (electric, magnetic, acceleration or gravitation field), see, e.g., \cite{yuak2001}.    
Optical BO and Zener tunneling of light waves have been recently observed in time-resolved experiments in 2D\cite{pertsch} and 1D\cite{sapienza} optical superlattices with refractive index gradient along the growth direction. Phonon BO
and Raman spectra of phononic WSL states were described in semiconductor multilayer solid structure based on acoustic-phonon cavities with different thicknesses\cite{lanzillotti}. More recently, WSL have been also observed in one-dimensional elastic systems\cite{Gutierrez}.

In this Letter we predict analytically and confirm experimentally the existence of acoustic Bloch oscillations in two-component ultrasonic superlattice made of layers of Methyl Methacrylate (Plexiglas) and water cavities.
The system of Plexiglas' layers, all having the same chosen thickness, separated by water cavities with a given gradient of thicknesses, is equivalent to a set of acoustic-wave cavities with a gradient of local resonant frequencies. 
In this system the water cavities act as resonant cavities, the resonant acoustic modes being well confined, which results in well-pronounced acoustic WSL in the transmission spectra through the structure. This in turn will result in time-resolved Bloch oscillations in the transmission (and reflection) spectrum for an incident acoustic  pulse with the proper spectral position and width. 
We will also show that for a particular gradient of water layer thicknesses in the superlattice, an overlap of two acoustic bands occurs which results in resonant enhancement of the phonon transmission through the superlattice due to resonant Zener-like effect of spatially-overlapping phononic states belonging to neighboring minibands.
\begin{figure}[ht]
\includegraphics[width=0.40\textwidth]{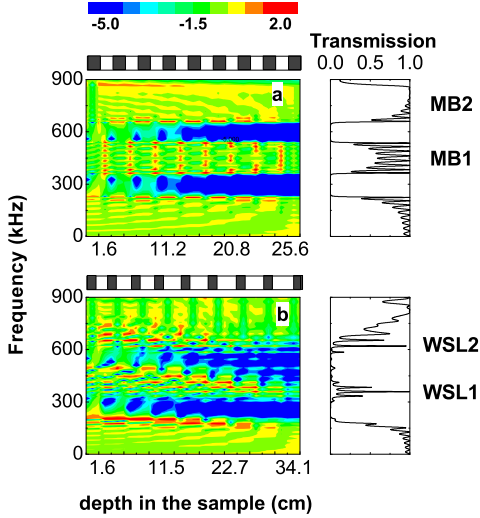}
\caption{(Color online) Transfer-matrix calculation of the intensity distribution (log$\left|P(z)\right|^2$) through a sonic superlattice made of eight coupled water cavities. (a) Flat band case, $\Delta (1/d_W)=$0$\%$, where two minibands (MB1 and MB2) are clearly seen. (b) Tilted band case, $\Delta (1/d_W)=$6$\%$, where Wannier-Stark like ladders (WSL1 and WSL2) are also shown.  The structure of coupled water cavities is schematically shown above each panel; dark regions represent Plexiglass layers. The right panels depict the transmission spectra across the total structure.}
\label{fig:Transmission}
\end{figure}

We start with the evolution equation for the Bloch wavevector $k_{Z}$ (along the axis of the periodic system):
 \begin{equation} 
  \dot{k}_{Z}= \frac{\partial k_{Z}}{\partial\omega}\frac{\partial\omega}{\partial Z}\dot{Z}
 =\frac{\partial\omega}{\partial Z}, 
\end{equation} 
where $\omega$ is phonon frequency and $\partial\omega /\partial k_{Z}=v_{Z}=\dot{Z}$ is the group velocity of the Bloch phonon wave packet.  
Equation (1) presents a generalization for phonons of the ``acceleration theorem'' for the evolution of electron Bloch wavevector ${\bf k}$ in an external electric field ${\bf E}$, $\hbar\dot{{\bf k}}=e{\bf E}={\bf F}$, with  $\partial\omega /\partial Z$ playing the role of the external ``wave'' force $F_{Z}/\hbar$. 

The dispersion equation for the longitudinal Bloch phonons (sound waves) propagating along the $z-$axis of two-component superlattice is the following, see, e.g., Ref. \cite{yuak1991}: 
 \begin{eqnarray}
&&\cos k_{Z}d=\cos(\frac{\omega d_{A}}{c_{A}})\cos(\frac{\omega d_{W}}{c_{W}})\nonumber \\
&&-\frac{1}{2}(\frac{\rho_{A}c_{A}}{\rho_{W}c_{W}}+\frac{\rho_{W}c_{W}}{\rho_{A}c_{A}})\sin(\frac{\omega d_{A}}{c_{A}})\sin(\frac{\omega d_{W}}{c_{W}}), 
\end{eqnarray}
where  $d_{A,W}$, $\rho_{A,W}$ and  $c_{A,W}$ are the thicknesses, densities, and sound velocities in the layers of elastic materials A and W, $d=d_{A}+d_{W}$ is the period of the superlattice. 
No mode conversion (from longitudinal to transversal) has been considered in this problem since the propagating sound wave is assumed that impinges the elastic layers normally to their surfaces. 

In order to separate the Fabry-Perot resonances of the transmission through the weakly-coupled elastic A and W layers, let us also assume that there is big acoustic mismatch between layers A and W; i.e., $\rho_{W}c_{W}\ll\rho_{A}c_{A}$. The corresponding unperturbed systems have their resonances centered at $\omega_A\approx n\pi\frac{c_A}{d_A}$ and  $\omega_W\approx m\pi\frac{c_W}{d_W}$, being $n$ and $m$ integers. >From Eq. (2) it is easy to get a narrow isolated tight-binding phonon miniband centered at the first ($m=1$) Fabry-Perot resonance in W layers by imposing the condition $c_W/d_W<c_A/d_a$.
 For example, for $c_{W}/d_{W}=c_{A}/2d_{A}$ one has (in the reduced Brillouin zone $-$$\pi$$<$$k_{Z}d$$<$$\pi$): 
\begin{equation} 
\omega= \pi\frac{c_{W}}{d_{W}}+2\frac{c_{W}}{d_{W}}\frac{\rho_{W}c_{W}}{\rho_{A}c_{A}}\cos k_{Z}d 
\end{equation}    

Now, according to Eqs. (1) and (3), for a constant phonon driving force $F_{Z}/\hbar$$=\partial\omega /\partial Z$ the phonon group velocity $\partial\omega /\partial k_{Z}$ will oscillate with the Bloch frequency $\omega_{B}$=$F_{Z}d/\hbar$=$d\partial\omega /\partial Z$, in a direct analogy to electronic Bloch oscillations in a biased SSL.  

To demonstrate the Wannier-Stark like ladders and the corresponding Bloch oscillations predicted by the analytical model above, we have considered a superlattice made by alternating fluid layers and layers of elastic material. In particular, experimental measurements have been performed on a set of multilayers (ML) samples consisting of 8 coupled water cavities W$_\ell$ enclosed by 9 layers of Plexiglas. 
Transfer-matrix (TM) calculations\cite{bre} of the sound intensity transmitted through this multilayer as a function of frequency have been performed by using the following data. As parameters we have taken $\rho_A=$1.19 g/cm$^3$, $c_A=$2.65$\times$10$^5$cm/s and $\rho_W=$1 g/cm$^{3}$, $c_W=$1.48$\times$10$^5$cm/s for the density and sound velocity in Plexiglas (layers A) and water layers, respectively. Viscosity effects are physically irrelevant to the studied phenomena and are not reported here.

First, a perfect superlattice consisting of alternating layers with equal thickness, $d_A=d_W=$0.16 cm, has been studied by TM simulations. 
The results for this case (see upper panels in Fig. 1) show that, though the contrast of acoustic impedances between Plexiglas and water is low, a group of modes MB1 is fairly well confined in the water cavities. The localization of these modes is visible as a set of elongated hot spots surrounded by dark regions representing the bandgaps. 
The group is centered at the first ($m=$1) Fabry-Perot resonance of water cavities and corresponds to a discrete analog of a miniband [see Eq. (3)]in our eight-cavity structure. The liner midfrequency of MB1 obtained from TM calculation is $\nu_1^{TM}=$446 kHz, which agrees with the first Fabry-Perot cavity resonance in the water cavities; i.e., $\nu_1\approx(1/2)(c_W/d_W)=$463 kHz. The localization of the states belonging to the group MB2 is not as clear as in MB1 because MB2 contains states that are also localized in the Plexiglass layers. 

\begin{figure}[ht]
\includegraphics[width=0.40\textwidth]{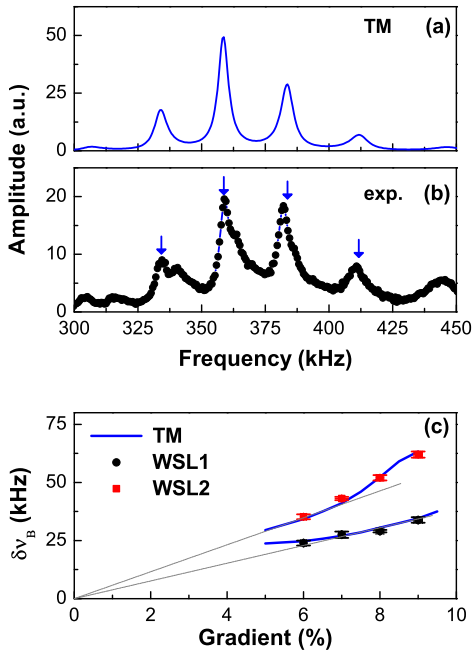}
\caption{(Color online) (a) Amplitude of transmission calculated by transfer matrix (TM) for the multilayer described in Fig. 1(b). The acoustical analogue of the electronic Wannier-Stark ladder is seen as a series of almost equidistant transmission peaks. (b) The corresponding experimental spectrum obtained from time-domain measurements in transmission. (c) Spacing between peaks ($\delta \nu_B$) in the corresponding WSL as a function of the gradient $\Delta (1/d_W)$. The gray dotted lines are guides for the eye and allow to distinguish the linear WSL regime.}
\label{fig:figure2}
\end{figure} 

In our acoustic ML structures the frequency of states localized in the water cavities depends approximately linearly on the driving force, $\partial\omega /\partial Z$, with the slope given by $md_W\partial\omega /\partial Z$ ($m=$1, 2, 3, $\ldots$), where $\partial\omega /\partial Z=\pi c_W\frac{\partial}{\partial Z}(1/d_W)$. 
For example, for the case $m=$1, a linear variation of frequency is obtained by introducing a constant variation of the inverse of the thicknesses of the water cavities $1/d_W$ [see Eq. (3)]:
\begin{equation} 
\delta \omega\equiv \omega(z_\ell)-\omega(z_{\ell-1})=\pi c_W\delta(\frac{1}{d_{W}}) 
\end{equation}
\noindent where $\delta (1/d_W)\equiv(1/d_{W_\ell}-1/d_{W_{\ell-1}})=const$, the subindex $\ell$ defines the ordering of cavities in the structure, $\ell=1,2,\ldots 8$.  From here onwards, the results will be given in terms of the dimensionless parameter $\Delta (1/d_W)\equiv[(1/d_{W_\ell})-(1/d_{W_{\ell-1}})]/(1/d_{W_1})$, where $d_{W_1}=$0.16 cm is the thickness of the first, $\ell=$1, cavity. Figure 1(b) shows the pressure intensity distribution for $\Delta (1/d_W)=$6$\%$.
Notice that the group MB1 is linearly tilted when a linear gradient is imposed over the cavity thicknesses.
It becomes in a discrete sequence of frequency levels, WSL1, strongly localized within one of the water cavities and one therefore obtains a well-defined spacing $\delta \omega_{B1}$ between levels.
 This set of equally spaced levels are the acoustical equivalent of the Wannier-Stark levels. 
 The spacing between levels being approximately determined by $\delta \omega_{B1}=\omega (z_1)\Delta (1/d_W)$.
  The discussion about the levels in MB2 is more complex, but a set WSL2 with level spacing $\delta \omega_{B2}$ can also be defined. In what follows we will focus our discussion in the set WSL1.

\begin{figure}[h]
\includegraphics[width=0.35\textwidth]{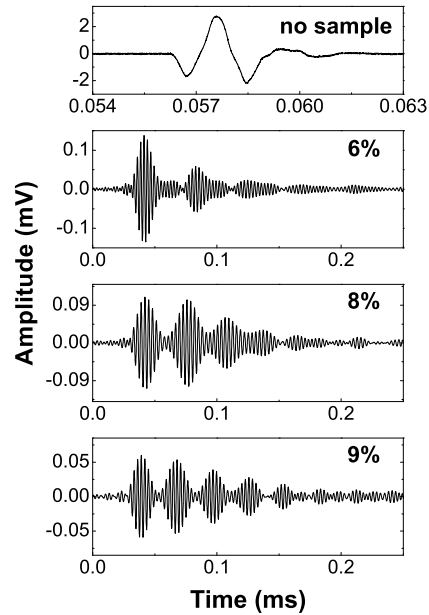}
\caption{Temporal response of the system for various values of the gradient. The top panel shows the unperturbed probe pulse without superlattice sample. The observed oscillations are the acoustical counterpart of the electronic Bloch oscillations. The period, $T_B$, measured are 41.5 $\mu$s, 32.7 $\mu$s, and 29.3 $\mu$s for gradients 6$\%$,8$\%$, and 9$\%$, respectively.}
\label{fig:TemporalResponse}
\end{figure}

Therefore, for the  gradient considered (6$\%$) the frequency separation (between states in WSL1) predicted by the simple analytical model previuosly described is $\delta \nu_{B1}\approx $27.8 kHz, which is close to that found between peaks in Fig. \ref{fig:figure2}(a), which shows the transmission spectrum calculated by TM for the case of a gaussian beam $f(\omega)$ centered at 500 kHz crossing the gradient superlattice. A system having such discrete sequence of frequency levels with level spacing $\omega_B$ is the acoustical equivalent of the electronic WSL, and is expected to exhibit acoustical Bloch oscillations of period $T_B=2\pi/\omega_B$ as it has been experimentally demonstrated.

Transmission measurements have been performed in a water tank by a simple experimental set up consisting of two transducers, an ultrasonic square wave pulser/receiver and a digital oscilloscope connected to a computer to store the data. 
In brief, the emitter transducer, which is put near (lees than 1 cm) the sample's surface, is employed to excite a longitudinal wave that crosses the structure and is recorded by a the receiver transducer, which is also placed near the last layer of Plexiglas. 
The central frequency of both transducers is 500 kHz, their bandwidth being approximately 50$\%$. 
The temporal resolution is 0.4 $\mu$s.
Effects of mode conversion in this system can be considered as negligible since the excitation and detection involve only vibrations normal to the interfaces and the elastic layers are very thin. 
 
\begin{figure}[ht]
\includegraphics[width=0.40\textwidth]{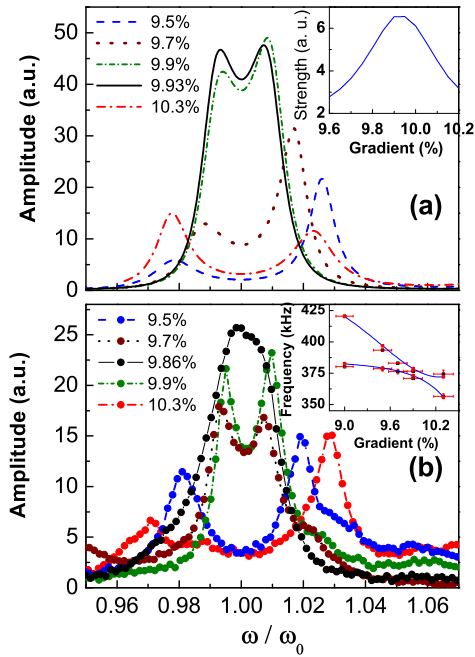}
\caption{(Color online) (a) Transfer matrix (TM) calculation of the transmission spectra of a tilted superlattice around the value of gradient, $\Delta (1/d_W)$), where the first anticrossing of the acoustical WSL and hence Zener-like resonant effect occurs. The amplitude is plotted as a function of the reduced frequency, where $\omega_0$ is the central frequency. The inset shows the strength (see text) of the resonant peak arising from the interaction between the two states located in WSL1 and WSL2, respectively [see Fig. 1(b)].(b) The transmission spectra measured by time-resolved transmission experiments. The inset shows a comparison between the behavior obtained by TM simulations (full lines) and experiments (symbols with error bars).}
\label{fig:Zener}
\end{figure}

Figure 2(b) shows, as a typical result, the experimental spectrum for the amplitude of states WSL1 corresponding to a gradient of 6$\%$. The blue arrows indicate the position of the peaks determined by TM simulation (see Fig. 2a). 
The frequency spacing between peaks is $\delta \nu^{exp}_{B1}\approx $24.3 kHz, which agrees with the value calculated by TM $\delta \nu^{TM}_{B1}\approx $24.8 kHz, and with the one obtained by the analytical model.
Also, an overall good agreement is obtained between theoretical and experimental spectra in the full range of frequencies of interest. 

The comparison between measured frequency spacings $\delta \nu_B$ as a function of the gradient with the ones calculated by TM is shown in Fig. \ref{fig:figure2}(c). 
It is remarkable the very good agreement between theoretical predictions and measured values. 
It must be pointed out that the true WSL regime is characterized by a linear dependence of $\delta \nu_{B}$ as a function of gradient; i.e., $\delta \nu_{B}\rightarrow 0$ when $\Delta (1/d_W)\rightarrow 0$. 
To check this behavior, two straight dotted lines are plotted in Fig. \ref{fig:figure2}(c) to allow the distinction between the non-linear Fabry-Perot regime.
 It can be concluded from Fig. \ref{fig:figure2}(c) that the linear regime is shorter for WSL2 than for WSL1.

The time resolved transmission experiments are performed by sending the short pulse depicted in the upper panel of Fig. \ref{fig:TemporalResponse}. The Bloch oscillations corresponding to previously described WSL1 are also shown in Fig. 3 for various values of the gradient. The oscillation period $T_B$ experimentally observed decreases while increasing $\Delta (1/d_W)$ as expected.
Our previous analysis [see Fig. 2(c)] concluded that above 6$\%$ gradient a true WSL is formed in our sample.
>From the time-resolved data we can observed that indeed oscillations occur in transmission. More than five period oscillations are observed in the transmitted intensity with a period that decreases as $\Delta (1/d_W)$ increases. 
In addition, as the gradient increases the intensity transmitted decreases, which can be understood from the increases tilt of the bandgap (see also Fig. 1). The experimental data are in very good agreement with the calculated dependence, which is not shown here but can be directly concluded from Fig. \ref{fig:figure2}(c) where $\delta \nu_B=1/T_B$

To examine the Zener-like resonant effect, we performed TM calculations for gradient larger enough to observe that one level from WSL1 and another from WSL2 lead to resonant transmission through the superlattice. Figure 4(a) depicts the results showing that an important feature of the interaction between levels is its anticrossing behavior. The strength of the interaction, which is defined as the integral of the transmission peak, is shown as a inset in Fig. \ref{fig:Zener}(a) and its maximum value is achieved when the separation between peaks is minimum; i.e. at the gradient of 9.93 $\%$. 
Figure 4(b) shows the transmission spectra obtained from time-resolved transmission experiments confirming the TM predictions. 
It is possible to observe how the two states start to overlap and the saddle like curvature transforms into a sharp resonance, where the separation between levels is difficult to appreciate mainly due to the precision of our experimental set up. The inset in Fig. 4(b) plots the comparison with the TM simulations, the error bars showing that the main source of error comes from the precision in the definition of gradients, which comes from the mechanical inaccuracies associated to the construction of water cavities with correct thicknesses.
  

In summary, acoustic Bloch oscillations and the acoustic analogue of the resonant Zener-tunneling have been demonstrated in composites consisting of water cavities surrounded by layers of Plexiglas (methyl methacrylate). This is the most simple system used to date to observe acoustic Bloch oscillations showing that the phenomenon is indeed universal and very robust. 
We hope that this demonstration stimulates future work to create acoustic devices based on these phenomena.  

\begin{acknowledgments}
Work partially supported by the Spanish Ministry of Science and Education (MEC) under Project No. TEC2004-03545. H. S.-A. and Yu.A. K. acknowledge grants paid by MEC, with refs. AP2005-1079 and SAB2004-0166, respectively. We thank D. Torrent for useful discussions, F. Cervera for technical help and F. Agull\'o-Rueda for his critical reading of the manuscript.
\end{acknowledgments}

\end{document}